\documentclass[twocolumn, aps, prb]{revtex4-1}

\usepackage{amssymb}
\usepackage{graphicx}
\usepackage{natbib}

\begin{document}

\title{Graphene hyperlens for terahertz radiation}

\author{Andrei Andryieuski}
\email{andra@fotonik.dtu.dk}

\author{Andrei V. Lavrinenko}
\affiliation{Department of Photonics Engineering, Technical University of Denmark, Kongens Lyngby, DK-2800, Denmark}

\author{Dmitry N. Chigrin}
\affiliation{Institute of High-Frequency and Communication Technology, University
of Wuppertal, Rainer-Gruenter-Str. 21 FE, Wuppertal, D-42119, Germany }

\begin{abstract}
We propose a graphene hyperlens for the terahertz (THz) range.
We employ and numerically examine a structured graphene-dielectric multilayered stack that is an analogue of a metallic wire medium.
As an example of the graphene hyperlens in action we demonstrate an
imaging of two point sources separated with distance
$\lambda_{0}/5$. An advantage of such a hyperlens as compared to a metallic
one is the tunability of its properties by changing the chemical potential of graphene. We also propose
a method to retrieve the hyperbolic dispersion, check the effective
medium approximation and retrieve the effective permittivity tensor.
\end{abstract}


\maketitle

Rapidly developing terahertz (THz) science and technology acquired
a great deal of attention in recent years due to an enormous
potential for spectroscopy, communication, defense and biomedical
imaging \cite{Tonouchi2007,Jepsen2011,Kleine-Ostmann2011}. The natural
diffraction limit, however, restricts the resolution of the standard
THz imaging systems to about a wavelength, which is relatively large
(300 $\mu$m in the free space at 1 THz). To overcome
this restriction one can use the scanning near-field THz microscopy
that allows for even submicrometer resolution in the scattering (apertureless)
configuration \cite{Kersting2005},but such technique is slow. Another solution is to use a metamaterial
lens with artificially engineered properties, for example, a negative
index lens \cite{Pendry2000} or hyperbolic-dispersion lens (hyperlens)
\cite{Jacob2006}. While the negative-index material lens is far from being
implemented into the imaging systems due to high losses and narrow
resonant frequency range, the hyperlens has been experimentally demonstrated
in the microwave \cite{Belov2006} and optical \cite{Liu2007} regimes.
The hyperlens is able to convert evanescent waves into propagating ones
and to magnify a subwavelength image so that it can be captured by
a standard imaging system, a microscope, for example.

A hyperlens usually consists of metal-dielectric layers (in ultraviolet
and optical ranges) or of metallic wires (infrared and microwave ranges).
Due to the employment of metal the properties of the hyperlens cannot
be tuned after fabrication. In contrast to metal, graphene, a two-dimentional material with striking electronic, mechanical and optical properties\cite{savage2012}, supports
surface plasmon polaritons in the THz range \cite{Jablan2009,Ju2011}
that are widely tunable by the change
of graphene's electrochemical potential via chemical doping,
magnetic field or electrostatic gating \cite{Chen2012}. Many plasmonic effects and photonic applications of graphene have been proposed \cite{Lee,Liu2011,Liu2012,Vakil2011,Tassin2012,Ju2011a}. Nevertheless, up to our knowledge, no graphene
based hyperlens for the THz range has been proposed so far (however,
there has been recently reported a graphene and boron nitride
hyperlens for the ultraviolet \cite{Wang2012}).

In this letter we propose to use structured graphene for the creation of a hyperlens in the THz range. To support our proposal
we investigate the effective properties of the hyperbolic graphene
wire medium and then construct a hyperlens out of it. We check numerically
the performance of a full-size three-dimensional (3D) and its homogenized two-dimensional (2D) analogue and
demonstrate that it has the desired subwavelength resolution and magnification.

Several requirements have to be satisfied for constructing the hyperlens
\cite{Jacob2007,Silveirinha2007}. First of all, an indefinite material (the permittivity tensor components
have opposite signs) with strong cylindrical anisotropy should be used. Namely the
radial permittivity $\varepsilon_{r}$ should be negative ($\varepsilon_{r}<0$) while
azimuthal permittivity $\varepsilon_{\theta}$ should be positive
($\varepsilon_{\theta}>0$). In this case the in-plane isofrequency
contour is hyperbolic:

\begin{equation} \label{Eq1}
\frac{q^{2}}{\varepsilon_{\theta}}+\frac{\kappa^{2}}{\varepsilon_{r}}=1,
\end{equation}

\noindent
where $q=k_{r}/k_{0}$ is the normalized radial wavevector component,
$\kappa=k_{\theta}/k_{0}$ is the normalized azimuthal wavevector and
$k_{0}=2\pi/\lambda_{0}=\omega/c$ is the wavenumber in vacuum. Then
the waves with $\kappa>1$, that are evanescent in vacuum, can propagate
in the hyperbolic medium. Mathematically this means that for every
$\kappa$ there exists a real-valued $q$. Moreover, the dependence $q(\kappa)$ should be as flat as possible.
That ensures the same phase velocities for all spatial components
(various $\kappa$). There are two possibilities for satisfying this requirement:
to select a material with either a large negative $\varepsilon_{r}$
or a small positive $\varepsilon_{\theta}$.

For high transmission propagation losses characterized by ($Im(q)$) should be as small as possible.
For the waves with $\kappa \ll 1$ the radial
wavevector reduces to $q\thickapprox\sqrt{\varepsilon_{\theta}}$,
so it is primarily $\varepsilon_{\theta}$ that determines losses. The incoupling and outcoupling of the waves to the ambient
medium should also be efficient. For normally incident waves
from a dielectric with a refractive index $n$ onto the flat interface
with a hyperbolic medium, the reflection coefficient is $R=\frac{n-q}{n+q}=\frac{n-\sqrt{\varepsilon_{\theta}}}{n+\sqrt{\varepsilon_{\theta}}}$,
so in order to minimize reflection one has to match the azimuthal permittivity
$\varepsilon_{\theta}$ with the permittivity $\varepsilon=n^{2}$ of
the surrounding medium. This requirement limits the range of $\varepsilon_{\theta}$. In addition to that to maximize the hyperlens transmission the Fabry-Perot
resonance condition should be satisfied

\begin{equation}\label{Eq2}
R_{2}-R_{1}=\frac{m\lambda_{\rm{eff}}}{2}=\frac{\pi m}{qk_{0}},
\end{equation}

\noindent
where $R_{1}$ and $R_{2}$ are the inner and outer hyperlens radius,
respectively, $\lambda_{\rm{eff}}$ is the effective wavelength and $m$
is an integer number. The ratio of the radii $M=R_{2}/R_{1}$ determines
the hyperlens magnification.

Finally, since no natural electromagnetic materials with a strong cylindrical
anisotropy exist, artificial effectively homogenous metamaterials have to be used. That means that its lateral geometrical
period $P$ should be much (at least 5-10 times) smaller than the period of the wave with the highest
$\kappa=\kappa_{max}$. So, for example, if we wish to construct the
hyperlens for the free-space wavelength $\lambda_{0}=50\mu$m that
supports the wave with the highest $\kappa_{max}=5$, then
the lateral period of the metamaterial should not be larger than $P_{\rm{max}}=\frac{1}{10}\frac{\lambda_{0}}{\kappa_{\rm{max}}}=1\mu$m.

\begin{figure}[htbp]
\includegraphics[width=8cm]{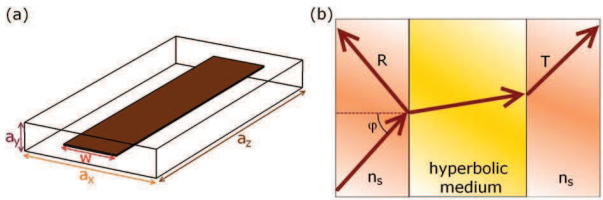}\caption{\label{Fig1}(a) The unit cell of the graphene wire medium consists of a graphene
stripe of the width $w$ embedded into a dielectric. (b) To characterize
the hyperbolic medium we calculated complex reflection $R$ and transmission
$T$ coefficients for various angles of incidence $\varphi$. A
block of the hyperbolic medium is placed between high-index $n_{s}$
dielectric layers.}
\end{figure}

First we analyzed the properties of the graphene wire medium itself. Its unit cell is
a rectangular block of dielectric ($\varepsilon_{D}=2.34$ corresponding to the low-loss polymer TOPAS) of the
size $a_{x}\times a_{y}\times a_{z}=0.2\times0.05\times1 \mu$m$^{3}$ ($a_{x}, a_{y} \ll P_{\rm{max}}$) with an embedded graphene stripe of the width $w$ depicted in Fig.\ref{Fig1}a.
We described graphene for the simulations in CST \cite{CST}
as a layer of thickness $\Delta=1$ nm with the permittivity $\varepsilon_{G}=\varepsilon_{D}+i\frac{\sigma_{S}}{\varepsilon_{0}\omega\Delta},$
where $\sigma_{S}$ is the surface conductivity of graphene.
\footnote{The surface conductivity
of graphene was calculated with the Kubo formula \cite{Hanson2008} 
in the random-phase approximation with the value of $\tau=10^{-13}$ s (which corresponds to rather conservative value of mobility $\mu=10^4 cm^2/(V\cdot s)$),
the temperature $T=300$ K and Fermi level $E_{F}=0.5$ eV. We compared the conductivity values that we used with the experimentally measured in the THz range \cite{Ren2012} and the relative difference was less than 7\%. Our test calculations for plasmons dispersion on a suspended graphene showed that numerical results differ from the analytical ones \cite{Falkovsky2008} less than 5\% for the selected effective thickness $\Delta=1$ nm.}

In order to retrieve the dispersion relation $q(\kappa)$ we simulated complex reflection $R$
and transmission $T$ coefficients for various angles of incidence $\varphi$ on a
hyperbolic medium slab (see Fig. \ref{Fig1}b) with the periodic
(unit cell) boundary conditions. We considered TM polarized waves (magnetic
field along the $y$-axis). The surrounding medium was a high refractive index $n_{S}$ dielectric. Then for each $\kappa$ and frequency $\omega$ we can restore $q$
\cite{Menzel2008}

\begin{equation}\label{Eq3}
q=\pm\frac{1}{k_{0}a_{z}}\arccos\frac{1-R^{2}+T^{2}}{2T}+\frac{2\pi m}{k_{0}a_{z}},
\end{equation}

\noindent
where $m$ is an integer number. Since we work in the long wavelength
limit, the challenging choice of the branch $m$ is not an issue,
it should be simply $m=0$. The choice of the sign should satisfy
the passivity condition $Im(q)>0$. Knowing the dispersion dependence $q(\omega,\kappa)$
we can restore the components of the permittivity tensor $\varepsilon_{r}$ and
$\varepsilon_{\theta}$ through the linear regression analysis of the dispersion equation (\ref{Eq1})

\begin{equation}\label{Eq4}
q^{2}=\varepsilon_{\theta}-\frac{\varepsilon_{\theta}}{\varepsilon_{r}}\kappa^{2}.
\end{equation}

The statistical coefficient of determination $R_{\rm{sq}}$ confirms (if
$R_{\rm{sq}}$ close to 1) the linear regression $q^{2}(\kappa^{2})$ and the homogenous approximation
validity. For the investigated graphene wire medium we observed $R_{\rm{sq}}>0.95$.
We should also note that this retrieval method is applicable not only to the hyperbolic medium, but to any metamaterial and that by selecting another polarization
and/or wave propagation direction it is possible to restore the whole
permittivity tensor.

\begin{figure}[htbp]
\includegraphics[width=8cm]{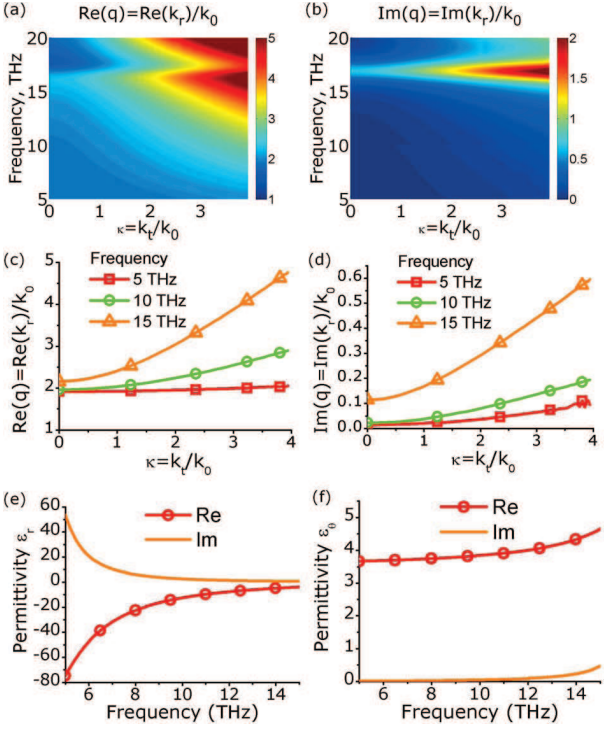}
\caption{\label{Fig2} Contour plots of restored (a) $Re[q(\omega,\kappa)]$ and (b) $Im[q(\omega,\kappa)]$
for a graphene stripe of $w=80$nm show absence of resonances at low
frequencies, but a resonance at $f=17$ THz. Looking at the $q(\kappa)$
for certain frequencies (5, 10, 15 THz) we observe the flat $Re[q(\kappa)]$
dependence (c) and smaller losses (d) for the frequency. Effective
radial permittivity $\varepsilon_{r}$(e) shows the Drude-like behavior,
while azimuthal permittivity $\varepsilon_{\theta}$ (f) is positive
with small losses.}
\end{figure}

An example of the restoration for the graphene stripe of width $w=80$ nm
is shown in Fig. \ref{Fig2}. The color contour graphs (Fig. \ref{Fig2}a,b)
show that $q(\omega,\kappa)$ is flat at low frequencies,
but exhibits a resonance around 17 THz. Detailed investigation of
the electromagnetic field behavior revealed a surface plasmon
resonance of the graphene stripe at this frequency. The $q(\kappa)$
isofrequency contours (Fig. \ref{Fig2}c,d) are flatter and
the losses are smaller at lower frequencies. Finally, the radial
permittivity $\varepsilon_{r}$ has the Drude-like dependence (Fig. \ref{Fig2}e)
with large negative values at the low frequencies, while azimuthal
$\varepsilon_{\theta}$ is positive and has small $Im(\varepsilon_{\theta})$
(Fig. \ref{Fig2}f). Thus it is advantageous to select a low operation frequency
for the hyperlens.

\begin{figure}[htbp]
\includegraphics[width=8cm]{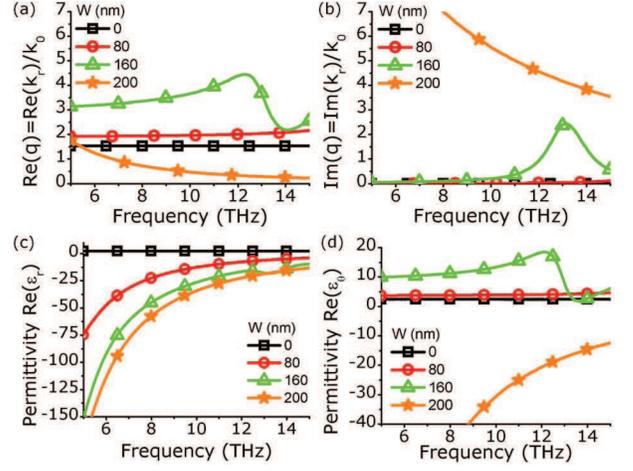}
\caption{\label{Fig3}Comparison of the properties of the graphene wire medium for various
stripe widths (0, 80, 160 and 200 nm). The radial wavevector shows larger
values of (a) $Re(q)$ (that means worse coupling to the surrounding
medium) and (b) $Im(q)$ (larger losses) for $w=160$ nm compared
to the width $w=80$ nm. Also, a "more metallic"
Drude behaviour of $\varepsilon_{r}$ (c) and higher azimuthal permittivity
$\varepsilon_{\theta}$ (d) nm is observed for $w=160$. The absence
of graphene ($w=0$) and full coverage ($w=200$ nm) show fully dielectric
and Drude-like behaviours, respectively.}
\end{figure}

In order to select the optimal geometrical design, we investigated
the dependence of the wire medium properties on the stripe width (see
Fig. \ref{Fig3}) starting from no graphene ($w=0$) to a full graphene
coverage ($w=200$ nm). As expected, in the absence of graphene we
restore a constant refractive index $n_{D}=1.53$ (Fig. \ref{Fig3}a)
with no losses (Fig. \ref{Fig3}b) and permittivities $\varepsilon_{r}=\varepsilon_{\theta}=n_{D}^{2}$
(Fig. \ref{Fig3}c,d), while for the full graphene coverage a typical
Drude metal-like behavior is observed for permittivities $\varepsilon_{r}=\varepsilon_{\theta}$.
Changing the width from $w=80$ nm, which we discussed above, to $w=160$ nm we observe
 larger values of $Re(q)$ for the normal propagation $\kappa=0$ (see Fig. \ref{Fig3}a) (and consequently worse coupling
efficiency), larger losses and red shift of
the resonance to $f=13$ THz (Fig. \ref{Fig3}b) and larger
negative permittivity $\varepsilon_{r}$ (Fig. \ref{Fig3}c). After
examining several widths we selected for the hyperlens demonstration the width $w=40$ nm (not shown in Fig. \ref{Fig3}) and the frequency 6 THz.

\begin{figure}[htbp]
\includegraphics[width=8cm]{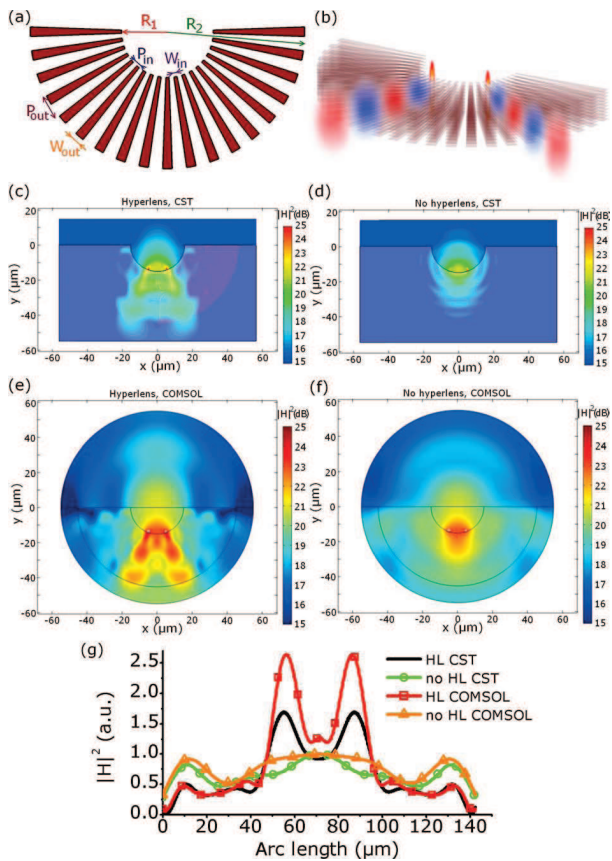}
\caption{\label{Fig4}(a) A single structured graphene layer that constitutes the hyperlens.
(b) An artistic 3D view of the hyperlens in action: the image of two
line sources is magnified with the hyperlens. Full size 3D CST simulation
of the hyperlens (c) in action and no hyperlens (d), when two magnetic
line sources separated by $\lambda_{0}/5=10\mu$m are emitting
TM polarized waves (magnetic field perpendicular to the plane of image). The CST results are in a good agreement with equivalent 2D COMSOL simulations of
the hyperlens (e) and no hyperlens
(f). Comparison of the intensity distribution at the output interface
of the hyperlens (g) confirms that the images are well resolved.}
\end{figure}

To check the suitability of the effective medium approach we simulated in the CST (time domain)
the full-size 3D hyperlens made of graphene stripes embedded into
dielectric ($n_{D}=1.53$). One
layer of the hyperlens is shown in the Fig. \ref{Fig4}(a). The input
and output periods, widths and radii were chosen as $P_{\rm{in}}=200$ nm,
$P_{\rm{out}}=600$ nm, $W_{\rm{in}}=40$ nm, $W_{\rm{out}}=120$ nm, $R_{\rm{in}}=15.12 \mu$m
and $R_{\rm{out}}=45.36 \mu$m, respectively. The radii are selected to satisfy the Fabry-Perot resonant condition (\ref{Eq2}). The layers of structured graphene
are assumed to be periodic in the direction perpendicular to the image
plane (period $a_{y}=50$ nm).
We should note that the specified sizes are realistic for fabrication. Multiple
graphene layers separated with a dielectric can be made up to the
size of 30 inches \cite{Bae2010}. Structuring of multiple graphene-dielectric
layers structure can be done with focused ion beam
milling or electron beam lithography.

Now we show the hyperlens in action when being excited with two sources (line magnetic currents) in vacuum separated
with distance $\delta=\lambda_{0}/5=10\mu$m (see the artistic 3D
view of the hyperlens in work in Fig. \ref{Fig4}b). In the presence of the hyperlens two
sources are well resolved at the output interface as two peaks separated
with $30\mu$m (Fig. \ref{Fig4}c) delivering the magnification $M=R_{2}/R_{1}=3$\footnote{We were limited by the computational power, so we took the hyperlens with small magnification $M=R_{2}/R_{1}=3$ (still, 12-core CPU with 48 Gb RAM executed the task in 3 days).}, while in case of the homogenous
dielectric cylinder (no graphene wires) we observe a single spot (Fig.
\ref{Fig4}d).

Then we compared the CST results with an equivalent 2D hyperlens simulation in COMSOL \cite{COMSOL} (scattering boundary conditions)
with homogenized permittivities $\varepsilon_{r}=-20.1+8.5i$, $\varepsilon_{\theta}=2.73+0.0029i$. The COMSOL
results with (Fig. \ref{Fig4}e) and without the hyperlens (Fig. \ref{Fig4}f) are in a good agreement with
the CST results. A comparison between them is shown in Fig. \ref{Fig4}g
where the wave intensity at the output interface of the lens is presented.
The intensity of the
peaks in the presence of the hyperlens is larger than in its absence, due to redistribution of the power. The intensity
simulated with the CST is smaller compared to COMSOL that is caused by
the coarser spatial discretization of the tapered wires with a staircase
numerical mesh in the CST. In both types of simulations the peaks are well resolved
according to the Rayleigh criterion. The 2D COMSOL simulation, however, took several minutes versus the 3-days long 3D CST modeling.

\begin{figure}[htbp]
\includegraphics[width=8cm]{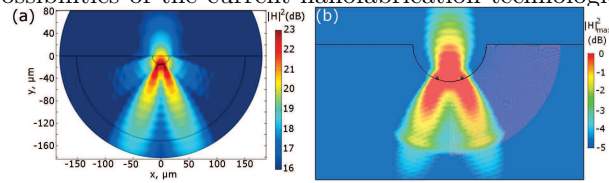}
\caption{\label{Fig5}(a) A thicker hyperlens with $R_{2}=10R_{1}$ magnifies two subwavelength
sources until the images can be captured with a conventional THz imaging system. (b)
Tracing the maximum of the broadband THz transient pulse shows that the
hyperlens works in the range of frequencies around 6 THz.}
\end{figure}

By making a hyperlens with larger radius $R_{2}$ one can achieve a
larger magnification. For example, selecting $R_{2}=10R_{1}$
gives the magnification $M=10$, so two point sources with separation
$\delta=10\mu$m are imaged to $100 \mu$m (see Fig. \ref{Fig5}a)
and then can be resolved with a conventional THz camera.

It is important to test the device performance under the pulse excitation. In the conventional THz time domain spectroscopy setup\cite{Jepsen2011} (THz-TDS), a very short (single cycle or even shorter) THz pulse is generated. Experimentally testing the hyperlens in the real THz-TDS would mean shining the short (and therefore broadband in frequency) transient pulse and then scanning with the THz near-field microscope and collecting the time-dependent signal at the output. We did a similar simulation in CST, exciting 2 sources with the Gaussian
pulse (central frequency $f_{c}=6$ THz, FWHM = 12 THz), recording the field with the time monitor and then imaging the maximal field in each point during the simulation time (Fig. \ref{Fig5}b). Since the graphene hyperlens is not based on a resonant
medium, it can operate in an extended range of frequencies and two sources
are well magnified and resolved (Fig. \ref{Fig5}b).

Due to reciprocity the hyperlens can be used not only for imaging, but also for THz power concentration into a small volume. We wish to note that the employment of metal for the considered hyperlens design is hardly possible. In order to obtain the same conductivity of the unit cell as of the regarded graphene stripes, the cross-section of the metallic wire (for example, silver \cite{Laman2008}) has to be of $2 \mu$m$^2$. Fabricating and arranging so thin and long metallic wires into the required pattern is beyond the possibilities of the current nanofabrication technologies.  We should emphasize that the scaling up the metallic wires together with the unit cell is not possible, since the period of the hyperlens should be subwavelength even for the higher-order spatial harmonics. Another important advantage of the graphene hyperlens compared to the metal based one is its tunability by the graphene chemical potential change. Thus it is possible to make the device reconfigurable and to resolve subwavelength features or concentrate THz pulses on demand. 

In conclusion, we have shown that structured graphene layers embedded
into dielectric (graphene wire medium) can be used to create a hyperlens. We have proposed the realistic
geometrical design for the hyperlens for the THz radiation and proved that it can resolve two line sources separated by a distance $\lambda_{0}/5$.
We also showed that time-consuming 3D simulations are in a good
agreement with the quick homogenized 2D hyperlens modeling, which
simplifies the hyperlens engineering.

\begin{acknowledgments}
The authors acknowledge A. Novitsky for useful discussions and M. Wubs for proof-reading.
A.A. acknowledges the financial support from the Danish Council for
Technical and Production Sciences through the GraTer (11-116991) project.
\end{acknowledgments}

\bibliographystyle{apsrev4-1}
\bibliography{AndryieuskiPRB2012}

\end{document}